# WHO IS AUTHORITATIVE? UNDERSTANDING REPUTATION MECHANISMS IN QUORA

Sharoda A. Paul[*], Lichan Hong[^], and Ed. H. Chi[^]

Palo Alto Research Center
3333 Coyote Hill Rd
Palo Alto, CA – 94304, USA.
paul@ge.com, lichan.hong@gmail.com, chi@acm.org

**ABSTRACT**

As social Q&A sites gain popularity, it is important to understand how users judge the authoritativeness of users and content, build reputation, and identify and promote high quality content. We conducted a study of emerging social Q&A site Quora. First, we describe user activity on Quora by analyzing data across 60 question topics and 3917 users. Then we provide a rich understanding of issues of authority, reputation, and quality from in-depth interviews with ten Quora users. Our results show that primary sources of information on Quora are judged authoritative. Also, users judge the reputation of other users based on their past contributions. Social voting helps users identify and promote good content but is prone to preferential attachment. Combining social voting with sophisticated algorithms for ranking content might enable users to better judge others' reputation and promote high quality content.

**INTRODUCTION**

People often augment their online information seeking with feedback from friends on general-purpose social networks and even strangers on specialized Q&A sites (like Yahoo! Answers). Quora is an emergent social-network based Q&A site, which aims at building a knowledge repository of questions and answers created and organized by the community. As of early 2011, Quora has been drawing 300,000 unique visitors a month and contains questions and answers on over 10,000 topics (KISSmetrics 2011). One of the challenges with social Q&A sites is to control the quality of the answers and to separate good answers from bad ones.

We were interested in understanding how the Quora community is building a high-quality knowledge repository. Specifically, we wanted to study the social mechanisms that help users judge the authoritativeness of answers, build reputation, and promote good answers, i.e., 1) How do users judge the authoritativeness of content and other users? 2) How do users build their own reputation and judge the reputation of other users? 3) How do users identify and promote high quality content?

We first studied the activities of Quora users by analyzing data across 60 question topics and 3917 users, collected through the site's publicly available RSS feed. Next, we conducted in-depth interviews of ten Quora users to gain insight into our research questions regarding authority and reputation mechanisms in Quora. We found that a prevalence of primary sources of information on Quora leads to users' perception that answers are high quality and authoritative. A social network based on real identities encourages users to build reputation while access to user histories enables them to judge the reputation of other users. Finally, the social voting system, combined with a unique ranking algorithm for answers, enables users to identify and promote high quality answers. Although, in general the crowd reaches consensus on good answers, popularity of users and preferential attachment (Barabási & Albert 1999) in voting can lead to the promotion of answers that might not necessarily be the best ones. These findings have important implications for the design of successful community knowledge-building sites.

We first discuss related literature on social Q&A. Next, we describe data collection methods followed by a description of site activity based on quantitative data. Then we provide deeper insights into research questions based on qualitative data. Finally, we discuss findings and implications.

**BACKGROUND**

Social Q&A refers to people asking questions online in a social context, either on social networking sites or using social search engines. Recently, several studies (Morris, Teevan & Panovich 2010; Paul, Hong & Chi 2011) have explored how users leverage general-purpose online social networks for information seeking. Morris, Teevan & Panovich (2010) conducted a survey study of how people ask questions using their status updates on Facebook and Twitter. They found that the majority of the questions were asking for recommendations and opinions, and pertained to technology and entertainment. Paul, Hong & Chi (2011) studied questions asked on Twitter and found that 42% of questions were rhetorical. Surprisingly, a considerable proportion (11%) of questions were related to health and personal issues like obesity, common ailments, relationships, and body-image. There was a low response rate (18%) on Twitter but answers were received quickly.

[*] Currently at GE Global Research, San Ramon, CA
[^] Currently at Google, Mountain View, CA

Another stream of research has studied community Q&A sites like Yahoo! Answers (Adamic et al. 2008), Live Q&A (Hseih & Counts 2009), AnswerBag and MetaFilter (Harper, Moy & Konstan 2009). Most studies of community Q&A have focused on characterizing Q&A sites or predicting response success and quality (Shah & Pomerantz 2010). Few have provided in-depth analysis of the social dynamics of user behavior or reputation mechanisms. Gazan (2011) provided such an in-depth study of how the re-design of AnswerBag led to violent withdrawal of users.

### Reputation and Authority in Social Q&A.

Two major challenges in crowd-sourced knowledge communities is controlling the quality of content, and providing users means to perform credibility judgments on content and users. Jurczyk & Agichtein (2007) stress the importance of understanding issues of authority and credibility in social Q&A sites. They proposed techniques to discover authoritative users for specific question categories by analyzing the link structure of the community. Tausczik & Pennebaker (2011) studied how offline and online reputation of contributors affect the perceived quality of contributions in MathOverflow, a social Q&A site dedicated to math. They found that both online and offline reputation were consistently and independently related to perceived quality of contributions. The credibility of content producers is often judged by the quality of their past contributions. Fiore, LeeTiernan, & Smith (2002) found that revealing author histories correlates with trust and an increased likelihood of reading authors' content.

There are two perspectives on the role of reputation in collaborative online communities like Wikipedia and Yahoo! Answers (Tausczik & Pennebaker 2011). One perspective is that user reputation should be minimized in order to encourage contributions from a wide base of users. The other is that user reputation should be made explicit to enable the community to identify and promote high quality contributions. When user reputations are made explicit through recommendation mechanisms, an important question is whether these recommendation mechanisms increase or decrease the phenomenon of preferential attachment. Jean-Samuel & Thomas (2009) found that an artist's offline reputation in the music industry is echoed within MySpace. Artists associated with major labels get most of the attention and tend to link to each other.

These studies indicate that it is important to understand how users build and judge reputation in online communities and the effect of reputation mechanism on perceived authoritativeness and quality of content is important.

### DESCRIPTION OF QUORA

On Quora, users ask and answer questions, and also comment on answers. They follow topics and other users. They can also assign topics to questions and create answer summaries. Further, Quora employees select users to be reviewers and admins. Admins have special privileges that allow them to review and delete content, lock questions, and flag suspicious activity.

Each question has it's own page which contains a ranked list of answers, topic tags assigned to the question, related questions, a list of users following the question, and past user activity on the question. Similarly, each topic has it's own page listing questions and answers on that topic, *open questions* that have not been answered, and the *best source questions* on that topic. Finally, each Quora user has a profile page that displays her recent activity, bio information, topics followed, and social graph.

### SITE ACTIVITY

We first wanted to understand the different kinds of activities users engage in. For this, we wanted to obtain a random sample of data on topics and users. Quora does not make site data available and does not allow crawling or scrapping the site[1]. However, each topic and user on Quora has a publicly available RSS feed. We created two data sets by capturing this RSS feed: *topics* dataset and *users* dataset.

The *topics* dataset contained all activity occurring in 60 randomly selected topics over a 27-day period in Apr-May 2011. Topics were selected as follows: for each question posted on the website (a list of all incoming questions are available at the questions log[2]), we randomly selected one topic that the question was tagged with. For 60 topics thus chosen, we captured the XML file of the RSS feed for the topic once every 10 minutes. We processed all XML files to remove redundant information and built a timeline of activity occurring in each topic for the given time period. For each topic, we captured questions asked on that topic, answers to question in that topic along with the names of the answerer and text of the answer. We could not capture asker names as questions on Quora don't have this information.

---

[1] Terms of Service: http://www.quora.com/about/tos
[2] Questions Log: http://www.quora.com/log/questions

The *users* dataset was created over 23 days immediately following the *topics* dataset collection. We selected all users (3917) appearing in the *topics* dataset, i.e. all users who had answered at least one question. For each user, we captured the XML file of their RSS feed once every 30 minutes and built a timeline of activities the user had participated in. For each user, we captured questions asked, answers added, topics followed, users followed, questions followed, answer voted on, and posts written.

**Topic Activity**

We collected data for a variety of topics such as music, movies, health and wellness, email, Judaism, and social analytics, etc. Table 1 shows the top five topics in our dataset.

| Top 5 topics | # of Qs | Top 5 answer topics | # of As |
|---|---|---|---|
| Survey Questions | 651 | Movies | 1642 |
| Movies | 509 | Music | 782 |
| History | 391 | History | 740 |
| Google | 259 | Psychology | 543 |
| Music | 255 | QTCA | 501 |

*Table 1. Top 5 topics by number of questions and answers.*

Over a 27-day period, 5317 questions were asked across 60 topics. Figure 1 shows the distribution of total number of questions and answers posted per topic. Per topic, more answers were posted (median = 66) than questions asked (median = 27).

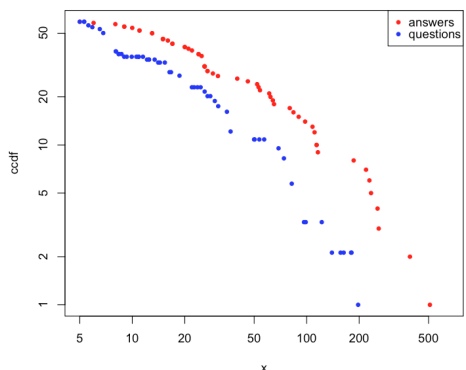

*Figure 1: Questions and answers per topic.*

2328 (43.8%) of new questions asked were answered within the 27-day period, with a total of 4908 answers received to these questions. The number of answers per question was low (median=1, avg = 2.1).

**User Activity**

Figure 2 shows the percentage of 3917 users who had performed different activities over 23 days:

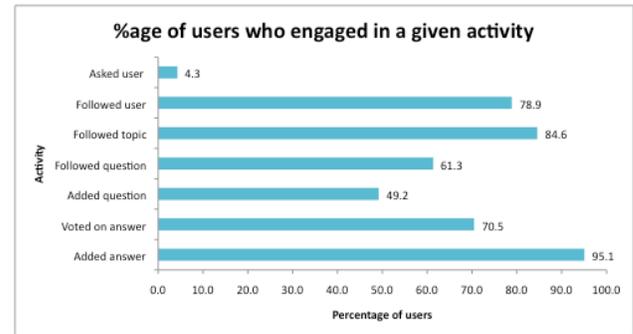

*Figure 2: Percentage of users who participated in different activities on Quora.*

The most popular activity was answering a question, followed by following topics and users. Interestingly, only 49.2% users participated in asking questions whereas 95.1% participated in answering questions. The activities of users had rough power law distribution, as seen in Figure 3. The distributions are similar to findings from other Q&A sites like Yahoo! Answers (Adamic et al. 2008) and StackOverflow (Mamykina et al. 2011) and suggests that we had indeed selected a random sample of users.

*Social voting*

Of the 3917 users, 71.4% had voted on at least one answer (median = 7 votes/user, average = 17.6 votes/user) and had together provided 49,327 votes, making it the most frequent activity on the site. Figure 3 shows that the distributions of votes per user (green line) had roughly a power law distribution.

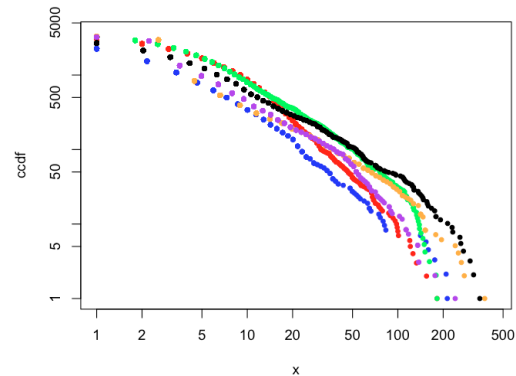

*Figure 3: User activities show rough power law distributions. Different colored plots are for the different activities from Figure 2.*

1902 (i.e., ~4%) of votes provided by users went to anonymous answerers. We analyzed the remaining answerers (9383) to explore how votes were distributed across answerers and found that this was again roughly a power law distribution (Figure 4) with mean = 5.1 votes, median = 1 vote. Thus, most answerers get few votes while a small number of answerers get lots of votes.

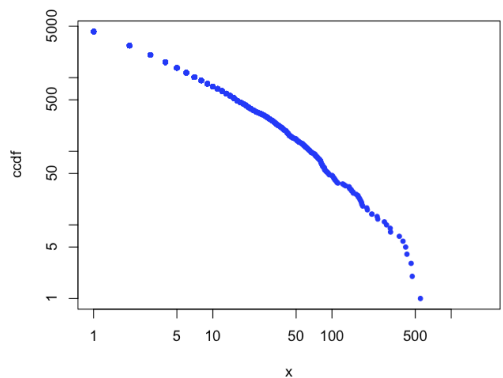

Figure 4: *Distribution of votes per answerer.*

## REPUTATION AND AUTHORITY IN QUORA

The quantitative data shows the wide variety of user activities on Quora (Figure 2). We wondered whether any of these activities were aimed at reputation building. For instance, 78.9% users followed other users and we wondered whether following users was partly for reputation building. We also wondered how users judge the reputation of other users before deciding to follow them. We also wondered whether votes were being used effectively to promote high quality content and whether votes were being used as a social signal to connect with popular users and build reputation. Finally, we saw that 95.1% users had participated in answering questions. We wondered whether writing answers was being used as a reputation building mechanism. Thus, the quantitative data led to many questions about how users were building reputation through different activities. To find answers to these questions, we conducted interviews with Quora users.

### User Interviews

We interviewed ten active Quora users. Interview participants were randomly sampled from the *users* dataset. We selected users with moderate to high activity levels since such users would potentially be familiar with different aspects of the site. We recruited participants by sending them a private message on Quora with a request to participate in a 30-minute interview. We compensated participants by entering them in a draw for a $50 Amazon gift card. Interview participants were 40% female and belonged to a range of professions such as technologists, librarians, journalists, politicians, and students. 8 participants were from the US (CA, CO, TX, OH) and the remaining from the UK.

Interviews were semi-structured and we asked participants questions about their perceived value of Quora and perceived difference with other Q&A sites; their motivations to produce content on Quora; their voting and following behavior; and generally about the reputation mechanisms on Quora.

Interviews were audio-recorded and transcribed. We used open coding (Strauss & Corbin 1990) to code the interview transcripts and iteratively developed our codes into interesting themes related to our research questions. We also read site content related to our research questions. Quora has several meta-level question topics about the product, the company, and user behavior and norms. Interview participants pointed some of this content to us while other content was found by searching the site. Next, we provide findings about authority and reputation in Quora as gathered from interviews.

*Primary sources provide authoritative answers*

Quora answers were deemed to be of high quality due to the presence of "primary sources of information". Primary sources of information (Yale 2008) provide first-hand testimony or direct evidence concerning a topic under investigation. Getting answers from primary sources of information on a question or topic was the founding vision of the site. The presence of primary source knowledge was critical to users' perception of the quality of answers and was one of the most valuable things about Quora. One user said:

"*What I really like about Quora, it's the primary source knowledge. For instance there was a question recently, 'Why is the Palo Alto City Hall so big?' and this morning I opened the site and found the Mayor of Palo Alto answering it… I feel like the primary sources on Quora are only growing… A lot of arts and humanities people are getting on there, and professors, and they really know their stuff. That's the value of Quora.*" -- P8

Not only were primary source answers considered to be high quality, but they were also considered authoritative. As another participant mentioned:

"*Quora has a much higher quality of people that answer. You get people who have actually been there and done that. For example, somebody asked the question 'Do bombs in real life have little digital count-down things like you see in the movies?' He got an answer from someone who used to be on the bomb squad in Singapore. So an actual bomb squad person answered…That's authoritative!*" -- P3

Questions that benefited from primary source knowledge were often factual questions that were hard to answer using search engines.

Several design and strategy decisions help Quora engage primary sources of information. Most importantly, the founders seeded the site with high quality contributors and primary sources. Additionally, several design features ensure that users get answers from primary sources and experts. The *Ask-to-Answer* feature let's users tag other Quora users in questions so that questions can be answered by the most appropriate answerers. This helps users get answers from those they consider authoritative or experts on the question. We found that 4% of 3917 users used this feature, and they together asked 219 users to answer questions. Finally, users felt that being able to assign topic tags to questions would enable the questions to be seen by the "right people".

### *Real identities lend credibility*

Unlike sites like Yahoo! Answers, users use their real identities on Quora due to the strict real-names policy. Further, Quora is designed to be a persistent social network based on these real identities. There has lately been significant debate about the value of real identities on social networking sites (boyd 2011). The proponents of real identity use on social networks feel that real identities ensure accountability and safety online. Quora users feel that real identities lend credibility to answers. One participant said:

"*The answers on other sites, you really have no idea who is answering it. You don't know who they are, so you don't know what their motives are, whereas on Quora you usually do.*" -- P9

Another participant stressed that having real identities helps users build reputation.

"*Because it's attached to your real name, Quora is more like a social networking site and a Q&A site, instead of being just a Q&A site. So there is some sense that you can build reputation by using the site and making connections in a way that you couldn't with Yahoo! Answers where everyone is anonymous.*" -- P10

This participant compared being on Yahoo! Answers to reading the anonymous comments section of a news story where there is "lots of trolling and unpleasantness". In contrast, having real identities on Quora made users accountable and hence Quora was a "pleasant place to be".

### *Building reputation is a motivator for participation*

Motivations for participation in Q&A sites can be either intrinsic (such as altruism or the desire to learn) or extrinsic (such as gaining social status or rewards) (Mamykina 2011). Several Q&A sites provide extrinsic motivators through explicit reward mechanisms, such as point systems or gamification mechanisms (Mamykina 2011). Quora does not have such points systems and hence it is interesting to study users' motivations.

Participants mentioned intrinsic motivators such as personal satisfaction in answering questions, satisfaction of curiosity, pleasure in researching new topics, and feeling competent and expert. Additionally, several extrinsic motivators were important for participants. Many participants mentioned that building reputation and building social capital were important motivators for participation on the site. For instance, one of our participants who works in politics said:

"*The way I justify to myself spending time there is that what you want to do is build connections with people and build reputation, especially if you work in politics, but in anything. Knowing people and having relationships with people is important to your life and livelihood; so you want to be perceived as somebody who is valuable. So that gets your answers to the top and people remember seeing your name.*" -- P10

Another participant, who is an entrepreneur and works in the technology industry in Silicon Valley, said that the most valuable aspect about Quora for him was networking and building reputation. He would deliberately answer questions to "maximize" his reputation, especially when he was new in Silicon Valley and wanted to gain the attention of "tech insiders". He said:

"*When there is someone I know that I would like to impress that has asked a question, I will make a special effort to answer in a high quality way*" -- P3.

In general, reputation building seemed to be especially important for participants who were located in Silicon Valley. This is rooted in the history of how the site started and the fact that Quora is heavily populated with the movers and shakers of Silicon Valley.

Recognizing and rewarding contributions through rating systems is often used by designers of social sites to encourage contribution. On Quora, social voting works as a reward mechanism. Users said that votes made them feel rewarded for their efforts and hence encouraged participation.

*"I feel Quora users are really discriminating about what they up-vote…being coherent and eloquent, and good grammar…are rewarded on Quora. My decent writing is rewarded and my research is rewarded. That makes me feel good."* -- P8

Also, votes from power users, such as the site's founders and Quora employees, encouraged users to participate. On the flip side, being down-voted by moderators was discouraging for participants.

### Judging other users' reputation

One of the important questions in crowdsourced knowledge-building communities like Quora is how users judge the reputation of others on the site. Specifically, are user reputations judged based on their real-world personas or are they judged based on online popularity and quality of contributions?

In Quora, each user can provide some bio information in their profile. Since Quora has a strict real-names policy that is enforced by the site, most users provide accurate information about their real world identities in their bios. This information helps others judge their authoritativeness on topics and question. When asked how she judged the expertise of other users on the site, one participant said:

*"Lots of times it's from what they have on their bio. So someone says they are an astrophysicist and answer questions about that topic, then likely they know what they are talking about."* -- P9

Also, users can view the history of other user's past interactions by visiting their user page. Participants mentioned that they peruse past history and follow users who have provided high quality answers or voted up interesting content. As one participant said:

*"I wouldn't follow [other users] on bio alone. I would follow them if they have some interesting activity on their feed that I agree with. Because whatever activity they have is going to appear on my feed so I want to make sure they are voting up stuff that I would have voted up too."* -- P8

Thus, most participants followed other users based on the quality of their Quora contributions. Our findings echo those from Yahoo! Answers where researcher applied machine learning techniques to identify which characteristics of askers and answerers determined quality of answers (Shah & Pomerantz 2010). They found that answerer's reputation as measured by points on their profile and the order of their answer in a list of answers were the most significant features for measuring answer quality.

### Social voting to build reputation

We wanted to understand whether voting was effective in helping users identify and reward good content. We asked participants whether they based their votes on the content of answers or the identity of the answerer? Did good answers generally get voted to the top of the answer list?

Participants said that they voted on answers that were well-written, seemed to be from an expert, and contained a point of view that they agreed with or that they wanted to promote. However, voting also had social functions. Participants mentioned using votes as social signals to draw the attention of influential people. For instance, one participant said:

*"I'm much more likely to vote up someone I know. That's more like a social signal because I know they'll get a notification in their feed and they'll see my name and be reminded of me."* -- P3

Another participant located in Silicon Valley said that people often voted to associate with well-known users in the start-up industry. However, most participants who were not located in Silicon Valley said that they rarely knew the users whose answers they voted up and did not care about the identity of the answerer when voting on answers. Thus, mostly users who were conscious about reputation building used voting as a social signal.

### Preferential attachment in voting

As Figure 4 indicates, few answerers receive a large number of votes while most answerers receive a small number of votes on their answers. This could indicate that popular users, or ones whose answers are often up-voted, attract more votes than ordinary users. Our interview data seemed to support this. Participants mentioned that generally good answers received votes, but people with more followers often had their answers voted up, even if they were not the best answers. Also answers with high number of votes attracted more votes, thus leading to the "rich-get-richer" phenomenon. One participant said:

*"People who have a lot of followers get their answers up-voted a lot, even if they aren't necessarily great answers… if the guy has a lot of followers, his answer will get up-voted even if it doesn't make any sense… Also for anything that has been up-voted a lot, it tends to attract more up-votes."* -- P10.

Participants mentioned that once an answer received a lot of up-votes and was ranked highly, it was hard

for other answers to get visibility, even if they added new or valuable information. As one participant said:

*"If you have a lot of up-votes, and you have a long answer, nobody sees any of the other answers. So that's the answer for that question on Quora, and it will be forever. If there is new information or the situation changes, that answers is going to be the first one that people see forever."* -- P10

There is concern among Quora users that 'clique-voting' is prevalent on the site (Quora 2011a). This refers to users endorsing answers as a group, even when those answers are dubious or bad. This would be the case where users voted on answers because friends or followees had voted on them. Participants said that they did not vote on answers because of friends. However, clique voting may not be deliberate on the users' part and may be an artifact of site design. Certain design features may be leading to clique voting. For instance, in their Newsfeed, users see the voting behavior of those that they follow. Thus, they are more likely to pay attention to answers that followees provide and vote on those answers. In future work if we have access to a complete dataset, we would like to analyze this quantitatively.

*PeopleRank for ranking answers to a question*
Due to these inherent shortcomings of social voting, merely using the number of votes to rank answers to a question might not be the best way to promote good content. Thus, answers to a given question are not ranked merely on the basis of vote count. Rather, a combination of metrics is used to rank answers to a question (Quora 2010). These metrics take into account explicit signals like votes received to the answer and quality of the answerer's past contributions. Up-votes from users who have written high quality answers in the past, help boost the rank of that answer. Also answers written by users who have written good answers in the past get ranked higher. Down-votes reduce the rank of an answer. Votes from those who have been "detected to have been gaming the system" are ignored. This system of determining a user's authority on a given question is called the PeopleRank (Quora 2010) of a user. Answers on a question are ranked according to the PeopleRank of the answerers.

There is concern (Quora 2011b) among users that PeopleRank might also be prone to the "rich-get-richer" phenomenon where users with high PeopleRank always draw more votes to their answers. Thus, though PeopleRank is designed to provide a more sophisticated mechanism for ranking answers than merely vote count, it is still prone to challenges in effectively signaling good content.

## DISCUSSION AND IMPLICATIONS

We conducted an in-depth study of reputation and authority in Quora. We described user behavior through an analysis of quantitative data related to user activity. We found that users participated in several activities that might be used to build reputation and to promote high quality content.

To answer our research questions, we conducted an interview study of Quora users. First, we wanted to understand how users judge the authoritativeness of content and other users. We found that content produced by primary sources of information was perceived to be authoritative. This suggests that encouraging participation from primary information sources is highly valuable. Primary sources do not pertain to CEOs and Mayors only, but include anyone who might have first hand information on a topic. For instance, participants suggested that getting more artists, professors, college graduates, librarians, and journalists on the site would lead to more authoritative content on those topics. Further, design features that help direct questions at experts ensure that questions will get authoritative answers.

The second question we studied was how users build reputation and judge the reputation of other users on the site. Building reputation is an important extrinsic motivator for producing good quality content. Users not only want to connect with others but also want to be deemed valuable to the community. When it comes to judging the reputation of other users, users look not only at bio information but also histories of past contributions. This is similar to past findings (Fiore, LeeTiernan & Smith 2002) about making user histories visible in social sites.

Finally, we explored how the crowd controls the quality of content on a social Q&A site, especially in the absence of explicit reward mechanisms like points systems. We found that in general the social voting system promotes good answers but, sometimes, popular users and answers get more votes than deserving answers. Similar results were seen for social voting on Digg (Lerman & Galstyan 2008) where users with larger social networks were more successful in getting their stories promoted to the front page. On Quora, the PeopleRank algorithm hopes to mitigate this by ranking answers based on a combination of several metrics. However, this leads to situations where users with high PeopleRank are always preferred in the ranking even when their answers are not the best.

Our findings have several implications for the design of social Q&A sites. Our first finding on primary sources suggests that we should encourage real identities and build social networks based on those identities. Real identities, and persistent social networks provide credibility and help users judge the authoritativeness of other users. Second, since users judge reputation based on others actions and past contributions, we should ensure that past actions and contributions are easily discoverable. This is especially important in the absence of sophisticated rewards systems that signal user reputation. Finally, simple social voting can help promote good answers in most cases but more sophisticated algorithms might be required to counter the effects of preferential attachment.

**FUTURE WORK AND CONCLUSIONS**

Though we gained deep insights into user behavior through qualitative data, future work remains on using quantitative measures to verify our qualitative findings. For example, could we measure whether primary source answers are higher quality than others by examining the average number of votes received to such answers as compared to other answers? Do popular Quora users get more votes to their answers as compared to regular Quora users? We could also quantitatively measure preferential attachment in voting. Finally, having access to the social graph of users would help us investigate how social voting patterns are affected by the reputation of users and the reciprocity of user relationships.

In summary, we studied reputation mechanisms in Quora in order to understand how users judge authoritativeness, build reputation, and promote high quality content. We examined user activities from quantitative data and then provided rich qualitative data to delve into issues of authority, reputation, and quality. Our findings provide important implications for designing community knowledge-building sites. In future work, and with a complete dataset from Quora, we can quantitatively examine our findings.